\documentclass[twocolumn]{aastex631}

\newcommand{\Ms}{M$_{\odot}$}
\newcommand{\Ls}{L$_{\odot}$}

\newcommand{\Ni}{$^{56}$Ni}

\newcommand{\mic}{$\mu$m}

\newcommand{\lmartinez}[1]{\textcolor{teal}{#1}} 

\shorttitle{Dust in SN~2005af }
\shortauthors{Sarangi et al.}

\begin{document}
\title{Two Decades of Dust Evolution in SN~2005af through {\it JWST}, \textit{Spitzer}, and Chemical Modeling }

\author[0000-0002-9820-679X]{Arkaprabha Sarangi}
\affiliation{Indian Institute of Astrophysics, 
100 Feet Rd, Koramangala, Bengaluru, Karnataka 560034, India} 
\affiliation{DARK, Niels Bohr Institute, University of Copenhagen, Jagtvej 155A, 2200 Copenhagen, Denmark}
\email{arkaprabha.sarangi@iiap.res.in}

\author[0000-0001-7473-4208]{Szanna Zs{\'i}ros}
\affiliation{Department of Experimental Physics, Institute of Physics, University of Szeged, D{\'o}m t{\'e}r 9, 6720 Szeged, Hungary}

\author[0000-0003-4610-1117]{Tam\'as Szalai}
\affiliation{Department of Experimental Physics, Institute of Physics, University of Szeged, D{\'o}m t{\'e}r 9, 6720 Szeged, Hungary}
\affiliation{MTA-ELTE Lend\"ulet "Momentum" Milky Way Research Group, Hungary}

\author[0000-0003-0766-2798]{Laureano Martinez}
\affiliation{Instituto de Astrofísica de La Plata (IALP), CCT-CONICET-UNLP, Paseo del Bosque S/N, B1900FWA, La Plata, Argentina}

\author[0000-0002-9301-5302]{Melissa Shahbandeh}
\affiliation{Space Telescope Science Institute, 3700 San Martin Drive, Baltimore, MD 21218, USA}

\author[0000-0003-2238-1572]{Ori D. Fox}
\affiliation{Space Telescope Science Institute, 3700 San Martin Drive, Baltimore, MD 21218, USA}

\author[0000-0001-9038-9950]{Schuyler~D.~Van~Dyk}
\affiliation{Caltech/IPAC, Mailcode 100-22, Pasadena, CA 91125 USA}

\author[0000-0003-3460-0103]{Alexei~V.~Filippenko}
\affiliation{Department of Astronomy, University of California, Berkeley, CA 94720-3411, USA}

\author[0000-0002-6991-0550]{Melina Cecilia Bersten}
\affiliation{Instituto de Astrofísica de La Plata (IALP), CCT-CONICET-UNLP, Paseo del Bosque S/N, B1900FWA, La Plata, Argentina} 
\affiliation{Facultad de Ciencias Astronómicas y Geofísicas Universidad Nacional de La Plata, Paseo del Bosque S/N B1900FWA, La Plata, Argentina}
\affiliation{Kavli IPMU (WPI), UTIAS, The University of Tokyo, Kashiwa, Chiba 277-8583, Japan}

\author[0000-0001-9419-6355]{Ilse De Looze}
\affiliation{Sterrenkundig Observatorium, Ghent University, Krijgslaan 281 – S9, 9000 Gent, Belgium}

\author[0000-0002-5221-7557]{Chris~Ashall}
\affiliation{Institute for Astronomy, University of Hawai'i at Manoa, 2680 Woodlawn Dr., Honolulu, HI 96822, USA }

\author[0000-0001-7380-3144]{Tea Temim}
\affiliation{Department of Astrophysical Sciences, Princeton University, 4 Ivy Lane, Princeton, NJ 08544, USA}

\author[0000-0001-5754-4007]{Jacob~E.~Jencson}	
\affiliation{IPAC, California Institute of Technology, 1200 E. California Blvd., Pasadena, CA 91125, USA}

\author[0000-0002-4410-5387]{Armin~Rest}
\affiliation{Space Telescope Science Institute, Baltimore, MD 21218, USA}
\affiliation{Department of Physics and Astronomy, The Johns Hopkins University, Baltimore, MD 21218, USA}

\author[0000-0002-0763-3885]{Dan Milisavljevic}
\affiliation{Purdue University, Department of Physics and Astronomy, 525 Northwestern Ave, West Lafayette, IN 47907, USA }

\author[0000-0003-0599-8407]{Luc~Dessart}	
\affiliation{Institut d'Astrophysique de Paris, CNRS-Sorbonne Universit\'e, 98 bis boulevard Arago, F-75014 Paris, France}

\author[0000-0001-8033-1181]{Eli Dwek}	
\affiliation{Observational Cosmology Lab, NASA Goddard Space Flight Center, Mail Code 665, Greenbelt, MD 20771, USA}

\author[0000-0001-5510-2424]{Nathan Smith}
\affiliation{Steward Observatory, University of Arizona, 933 N. Cherry St, Tucson, AZ 85721, USA}

\author[0000-0002-1481-4676]{Samaporn~Tinyanont}
\affiliation{National Astronomical Research Institute of Thailand, 260 Moo 4, Donkaew, Maerim, Chiang Mai, 50180, Thailand}

\author[0000-0001-5955-2502]{Thomas~G.~Brink}
\affiliation{Department of Astronomy, University of California, Berkeley, CA 94720-3411, USA}

\author[0000-0002-2636-6508]{WeiKang~Zheng}
\affiliation{Department of Astronomy, University of California, Berkeley, CA 94720-3411, USA} 
\affiliation{Bengier-Winslow-Eustace Specialist in Astronomy} 

\author[0000-0002-0141-7436]{Geoffrey C. Clayton}
\affiliation{Department of Physics \& Astronomy, Louisiana State University, Baton Rouge, LA, 70803 USA}

\author[0000-0003-0123-0062]{Jennifer Andrews}
\affiliation{Gemini Observatory, 670 N. Aohoku Place, Hilo, Hawaii, 96720, USA }

\begin{abstract}
The evolution of dust in core-collapse supernovae (SNe), in general, is poorly constrained owing to a lack of infrared observations after a few years from explosion. Most theories of dust formation in SNe heavily rely only on SN~1987A. In the last two years, the \textit{James Webb Space Telescope (JWST)} has enabled us to probe the dust evolution in decades-old SNe, such as SN~2004et, SN~2005ip, and SN~1980K. In this paper, we present two decades of dust evolution in SN~2005af, combining early-time infrared observations with \textit{Spitzer Space Telescope} and recent detections by {\it JWST}. We have used a chemical kinetic model of dust synthesis in SN ejecta to develop a template of dust evolution in SN~2005af. Moreover, using this approach, for the first time, we have separately quantified the dust formed in the pre-explosion wind that survived after the explosion, and the dust formed in the metal-rich SN ejecta post-explosion. We report that in SN~2005af, predominantly carbon-rich dust formed in the ejecta, with a total mass of at least 0.02 \Ms. In the circumstellar medium, the surviving oxygen-rich dust amounts to about (3–6)$\times$10$^{-3}$ \Ms, yielding a total dust mass of at least 0.025 \Ms.  

\end{abstract}


\section{Introduction} \label{sec:intro}

Core-collapse supernovae (SNe) are considered to be significant dust producers in galaxies \citep{Szalai13, bou93, dwek_2006, matsuura2017, sarangi2018book, woo93, szalai_2019, gal11}. Dust produced in various SNe may vary in total mass, composition, and formation timescale. Infrared (IR) observations of several SNe in their first few years post-explosion have confirmed the presence of dust in their environment \citep[e.g.,][]{sug06, kotak_2006, and10, mei11, Szalai13, szalai_2019}. With the arrival of the \textit{James Webb Space Telescope (JWST)}, mid-infrared (mid-IR) observations uncovered cold dust reservoirs in several historic SNe such as SN~2004et, SN~1980K, SN~1993J, SN~2005ip, and SN~2017eaw  \citep{shahbandeh_2023, shahbandeh_2024, szanna_2024, szalai_2025}. 

The mass of dust inferred in an SN may vary from 10$^{-4}$ \Ms\ to as large as 1 \Ms\ \citep{wesson_2021, gal14, matsuura2017, priestley_2020, maria_2021, szalai_2019}. This variation is often correlated with the epoch of measurement and, to some extent, the instrument used for the detection. A moderate rate of dust formation is proposed to continue for several decades \citep{wesson2015, wesson_2021}, gradually building up the dust mass over that period. Alternatively, the effect of large optical depths in SN ejecta at early times is also proposed as a reason behind the variance of dust masses \citep{dwek_2019, sarangi_2022b} at various epochs of detection. Before the advent of {\it JWST}, there were hardly any supernovae other than SN~1987A, which was bright enough for any mid-IR instrument after three or four years post-explosion. The dust formation sequence remained uncertain due to the lack of late-time measurements. Our JWST Cycle 1 programs GO-2666 and GO-1860 (PI O. Fox) have achieved phenomenal success in detecting several SNe, which are already a few decades old. This enables us to account for the dust masses in a population of nearby SNe, deriving a trend on dust formation in SN in general. 

Dust that is detected in an SN could be formed prior to the explosion, in the mass-loss winds \citep{nozawa_2014, cherchneff2013b, verhoelst_2009}. Alternatively, it can form post-explosion, in the metal-rich ejecta and/or in a region of interaction between the SN shock wave and the circumstellar medium (CSM). Given the large distances of most SNe, observations do not have enough spatial resolution to locate where the dust is present or formed. Only nearby SN~1987A in LMC is an exception in this case, where the mass and location of dust have been traced over the last 35 years \citep{bou93, dwe15, mat11, matsuura_2015, arendt_2020, jones_2023}. Overall, for SNe in general, the observed IR spectra do not provide a strong constraint on the location or timescale of dust formation. In this context, the uncertainties for the estimated dust masses, the chemical type of dust, and the timescale for dust formation remain unresolved.

In this paper, we assemble the data obtained for SN~2005af at optical, near-IR, and mid-IR wavelengths, using archival data of Carnegie Supernova Project (CSP-I, optical photometry and spectrophotometry, \citealt{humay_2006,anderson_2024}), the \textit{Spitzer Space Telescope} (hereafter {\it Spitzer}), and the newly obtained photometric data from {\it JWST}. We have conducted a complete study of dust evolution over two decades, owing to our successful imaging of SN~2005af after $\sim 19$~yr post-explosion, through our {\it JWST} GO-2666 program. 

SN~2005af in NGC 4945, discovered in February 2005, is one of the closest known extragalactic SNe. The distance of the galaxy is debated to be either 3.47 $\pm$ 0.12 Mpc \citep{jacobs_2009, anand_2021} or 3.9~Mpc \citep{jacques_2005, filippenko_2005, kotak_2006}. The SN was classified as a Type II-P core-collapse SN whose light curve was powered by radioactivity, with negligible evidence of interaction with any CSM \citep{gutierrez_2017, anderson_2024}. The epoch of explosion has some uncertainty between MJD 53320.8 \citep{gutierrez_2017} and 53379 \citep{Szalai13}. We have chosen to use MJD 53320.8 based on more recent analysis of the SN light curve. Therefore, the epochs previously noted by \cite{Szalai13} and \cite{kotak_2006} are now offset by about 58 days.    

A general model of dust formation in SNe has been developed previously using a chemical kinetic approach \citep{sarangi2018book}. In this study, we used the specific ejecta properties that match the optical data of SN~2005af to derive its dust properties. Using that as a reference, we analyze the IR data to distinguish between the dust formed in a pre-explosion wind and post-explosion ejecta. This is the first study where we correlate  {\it JWST} and \textit{Spitzer} observations with dust-formation models to derive a consistent dust-evolution scenario in an SN.


\section{Rationale of the study}
\label{sec_rationale}

Here we report the first comprehensive analysis of an SN that connects early- and late-time mid-IR data with unique dust-formation models. We investigate the properties of the dust formed in SN~2005af over $\sim 2$ decades post-explosion, which is reflected in the detections with the {\it JWST} MIRI imager. This method enables us to distinguish between the various dust components and their locations, alongside estimating the timing of dust formation. 

In our recent surveys of dusty SNe such as SN~2004et, SN~1980K or SN~1993J \citep{shahbandeh_2023, szanna_2024, szalai_2025}, we had stressed on the large degeneracies in the location of dust and their chemical compositions, which cannot be uniquely resolved from the analysis of the IR data. In this paper on SN~2005af, we aim to connect the dust evolution in the ejecta and the surrounding medium from the time of explosion through the next two decades. When analyzing the IR data from {\it Spitzer} and {\it JWST}, the degeneracies on where the dust is located, and if the dust is likely to be silicate-rich or carbon-rich, were resolved with support from the chemical models of dust synthesis based on SN~2005af. In this regard, we shall first derive the theoretical estimates of dust formation in SN~2005af, so we have some references to use in our IR analysis. 

The paper is organized as follows. In Section \ref{sec_model}, we outline the model of dust formation in SN~2005af, based on SN properties derived from light curve modeling. Thereafter, in Sections \ref{sec_optical} and \ref{sec_Spitzer}, we summarize the optical data from CSP-I and the Keck telescope, and the available IR observations using \textit{Spitzer}. In section \ref{section_MIRI} we discuss the findings of our newly obtained {\it JWST} photometric data. In Section \ref{sec_Spitzer_model}, we describe the evolution of dust in SN~2005af from 2005 until 2024, which is consistent with all the observations. Finally, in Section \ref{sec_Discussion}, we present our findings about the overall physical background of the SN.

\section{Chemical Model of the Ejecta}
\label{sec_model}

\begin{figure*}
\vspace*{0.3cm}
\centering

\includegraphics[width=2.1in]{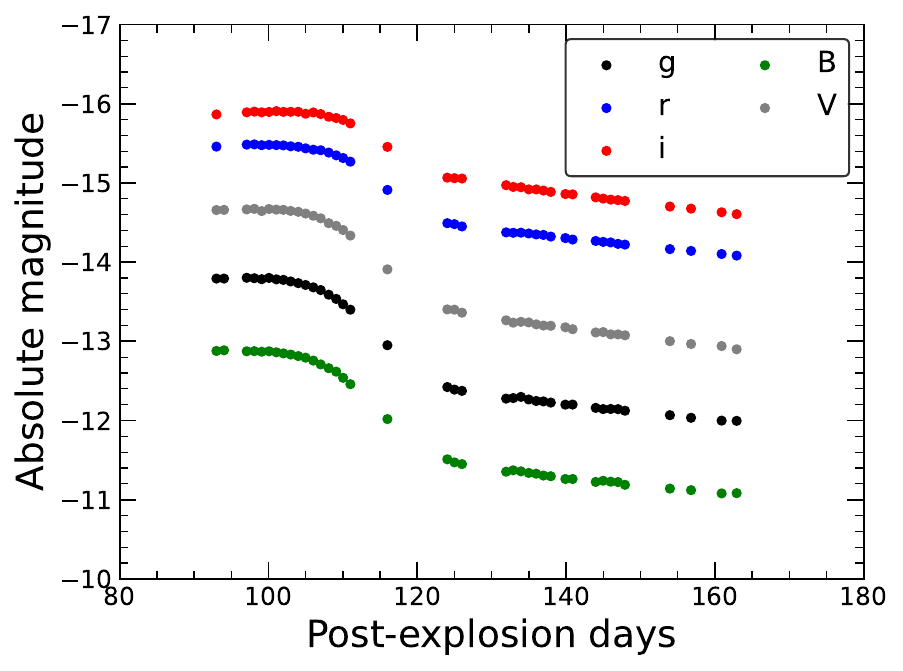}
\includegraphics[width=2.1in]{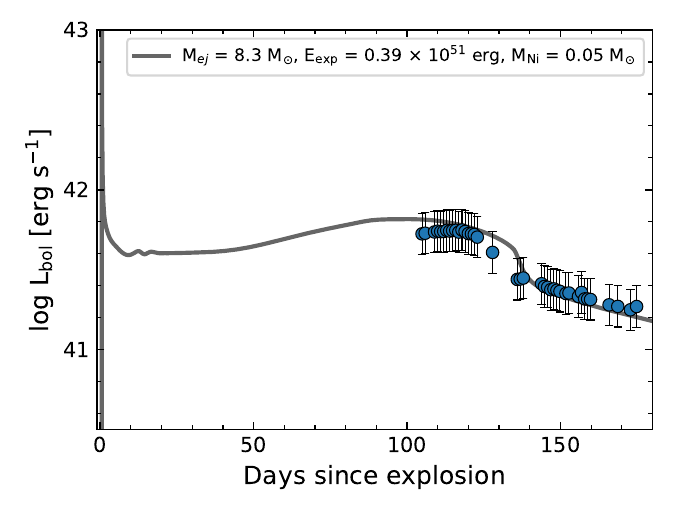}
\includegraphics[width=2.1in]{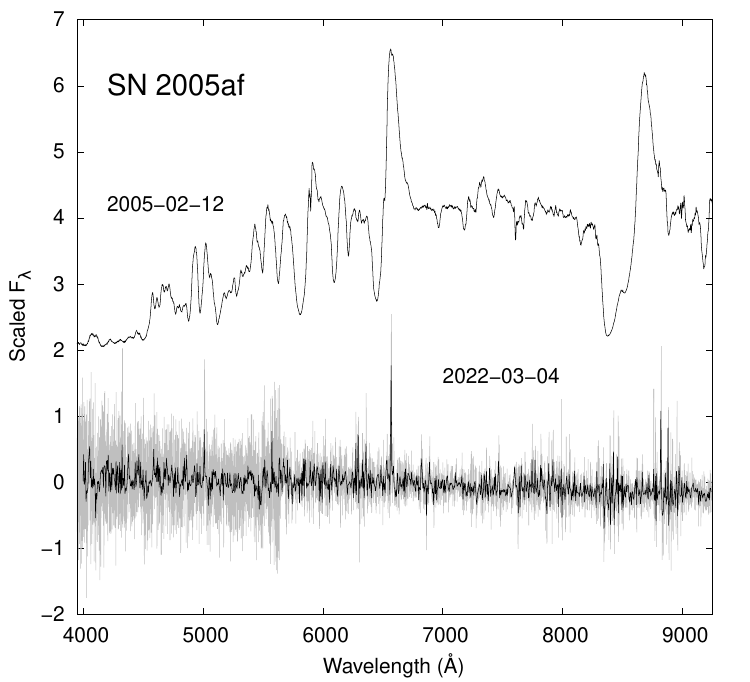}
\caption{\label{fig_lightcurve}\footnotesize{\textit{Left}: Fluxes for the \textit{B} (0.435 \mic), \textit{g} (0.477 \mic), \textit{V} (0.538 \mic), \textit{r} (0.622 \mic), and \textit{i} (0.761 \mic) bands of SN~2005af are presented for the epochs (day 92 to day 163 post-explosion) covered by the CSP-I \citep{anderson_2024}. \textit{Middle}: The best-fit light curve model, where the parameters are given in Table \ref{tab:table_lightcurve}.} \textit{Right}: Two high-resolution optical spectra by Keck/LRIS obtained at day 92 post-explosion, and also at day 6322 post-explosion.}   
\end{figure*}

The timescale, composition, and amount of dust produced in an SN depend on the evolution of physical conditions in the ejecta \citep{sarangi2018book, sarangi_2022b}. Specifically, the explosion energy, the mass of the ejecta, the chemical abundances, the velocity of the expanding gases, the mass of \Ni\ produced, and the clumpiness of the ejecta dictate the scenario of dust formation in SNe. 

Modeling the light curve of SN~2005af (Figure \ref{fig_lightcurve}), these physical parameters of the ejecta have been estimated.
Synthetic light curves were calculated using a one-dimensional (1D) hydrodynamical code that simulates the explosion of the SN and produces bolometric light curves \citep{bersten+2011}. 
This code has been widely used in the literature to model SN~II light curves and derive physical properties \citep[e.g.,][]{martinez+2020,martinez2_2022}.
The stellar models from \citet{sukhbold_2016} computed with the KEPLER code \citep{weaver+1978,woosley+2002} were used as pre-SN conditions to initialize the explosion.
Then, we estimated the progenitor and explosion properties of SN~2005af by comparing models constructed from different physical properties with the SN bolometric light curve. Therefore, we used the bolometric light curve of SN~2005af previously calculated by \citet{martinez1_2022} for the comparison.
The estimated physical parameters are reported in Table~\ref{tab:table_lightcurve}.
Using these results, we follow the prescription given by \cite{sarangi_2022b} to derive the dust masses as a function of time.

Based on the light-curve properties, SN~2005af is proposed to have a relatively low-mass progenitor, with a main-sequence mass of about 10 \Ms. This is the first model of dust formation applied to a core-collapse SN originating from such a low-mass progenitor. Previously derived dust formation models were mostly based on SN~1987A \citep{sluder2018, sarangi_2022b}, which is not ideal for this scenario. 

\begin{figure*}
\vspace*{0.3cm}
\centering
\includegraphics[width=2.2in]{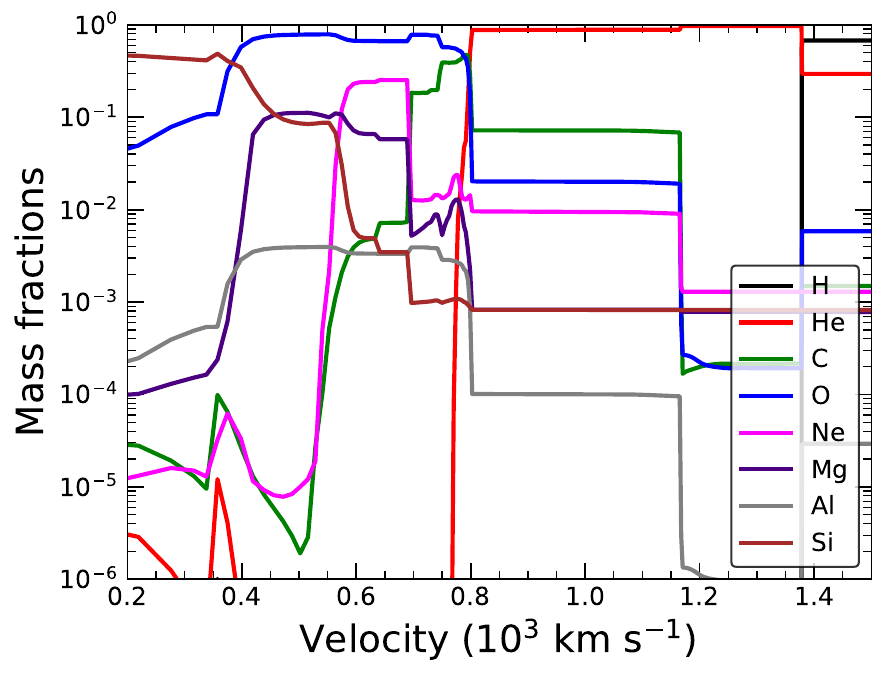}
\includegraphics[width=2.2in]{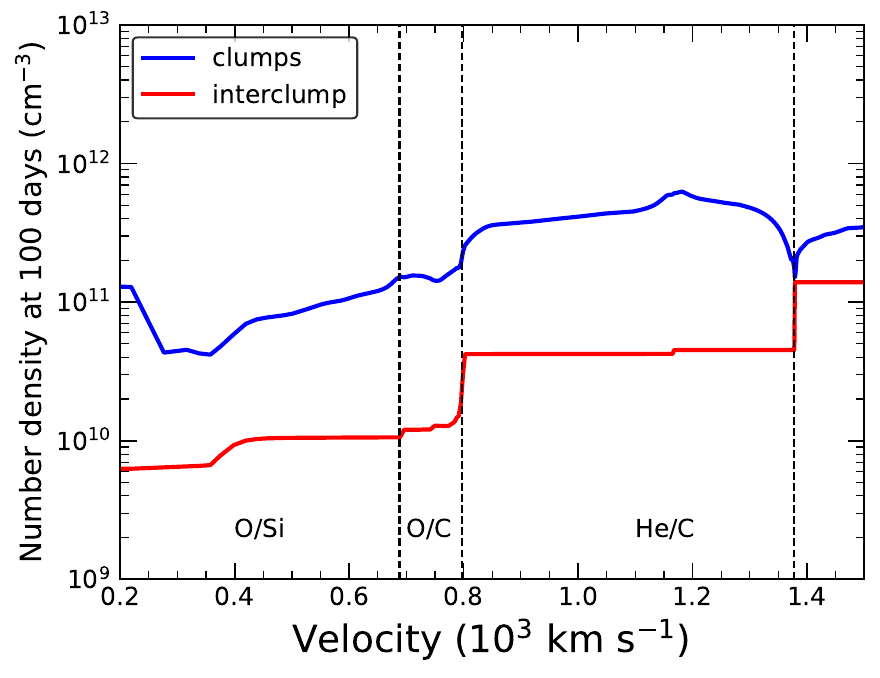}
\includegraphics[width=2.2in]{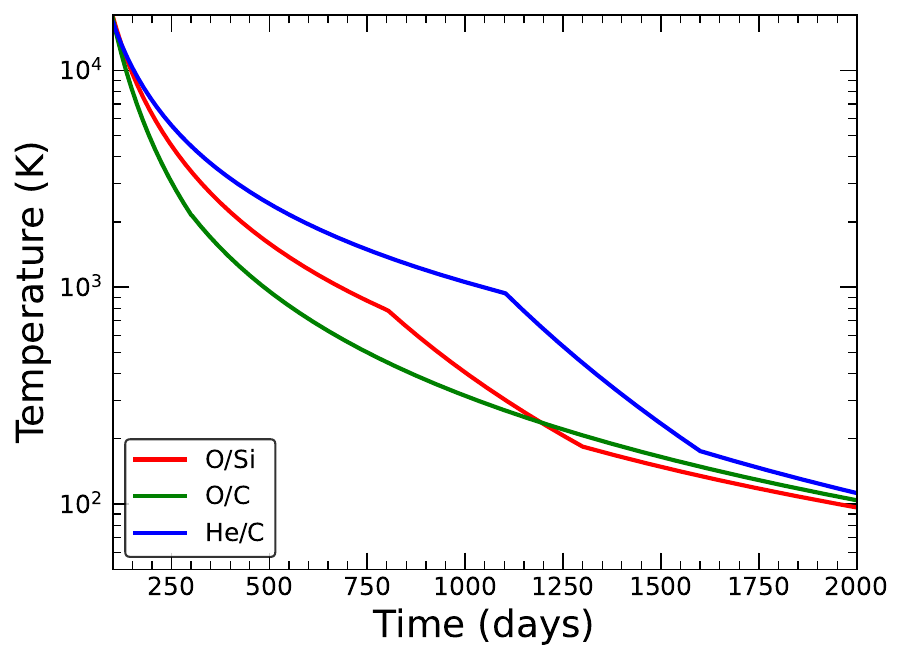}
\caption{\label{fig_initials}\footnotesize{The input conditions for dust-formation modeling in SN~2005af are presented. Elemental abundances (\textit{left}), gas densities at day 100 (\textit{center}), and gas temperatures (\textit{right}) are derived using the formalism adopted by \cite{sarangi_2022b} and the SN~2005af \lmartinez{ejecta} parameters derived in Section \ref{sec_model} (see Table \ref{tab:table_lightcurve}). The abundances and densities are shown in velocity space. The evolution of gas temperature is shown as a function of time for the O/Si, O/C, and He/C layers.  }}   
\end{figure*}

The ejecta is assumed to consist of an H-rich envelope surrounding an outer H-rich layer and an inner He core. Following the dynamics of SN ejecta (for details, see \citealt{sarangi_2022b}) given by \cite{tru99}, we find the ejecta core to have a velocity of about 2400 km s$^{-1}$ in this case. The metal-rich part is concentrated in the He core, which is confined within 1400 km s$^{-1}$. This He-core velocity is considerably slower than models based on SN~1987A presented by \cite{sarangi_2022b}, based on a 20~\Ms\ star. Slower expansion velocities correspond to more compact ejecta, with higher densities. The distribution of elements in the SN ejecta is shown in Figure \ref{fig_initials} in velocity space. Based on the abundances, the ejecta are stratified into O/Si, O/C, and He/C layers \citep{sar13}. The densities (Figure \ref{fig_initials}) for the clumpy medium are obtained from the light-curve analysis, while the densities of the interclump medium follow the homogeneous density structure given by \cite{tru99}. As in \cite{sarangi_2022b}, we consider dust formation only in the clumps. The evolution of the temperature, shown in Figure \ref{fig_initials}, also vary between the layers (O/Si, O/C, and He/C), with reference to the description by \cite{sarangi_2022b}.

Dust formation in the ejecta of SNe occurs through simultaneous phases of nucleation and condensation. We address the formation of dust in the ejecta of SN~2005af using a nonequilibrium chemical kinetic approach, which was developed in previous studies to account for any core-collapse SNe in general \citep{sar13, sar15, sarangi2018book, sarangi_2022b}. The detailed formalism described by \cite{sarangi_2022b} is used as the framework in this paper to derive the rate of dust formation in SN~2005af and its distribution inside the ejecta. We consider O-rich dust  (silicates and alumina) and C-rich dust (amorphous carbon and silicon carbide).

\begin{figure*}
\vspace*{0.3cm}
\centering
\includegraphics[width=6.5in]{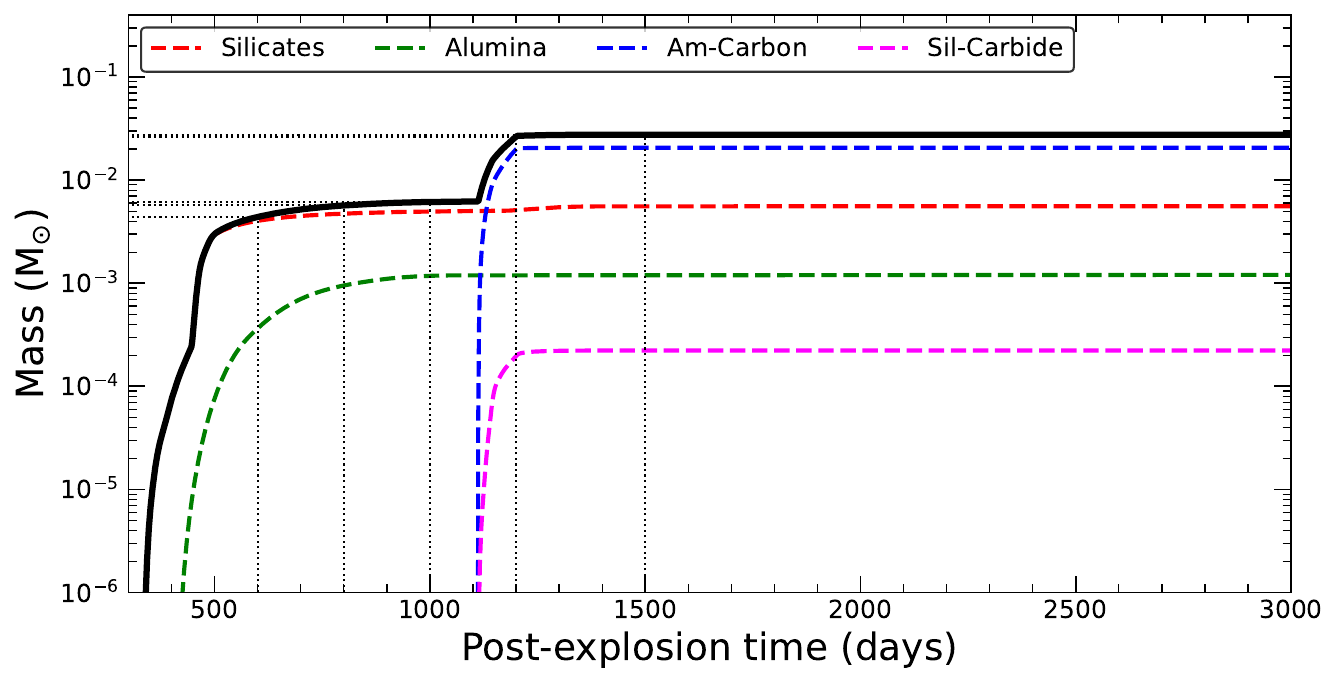}
\includegraphics[width=3.5in]{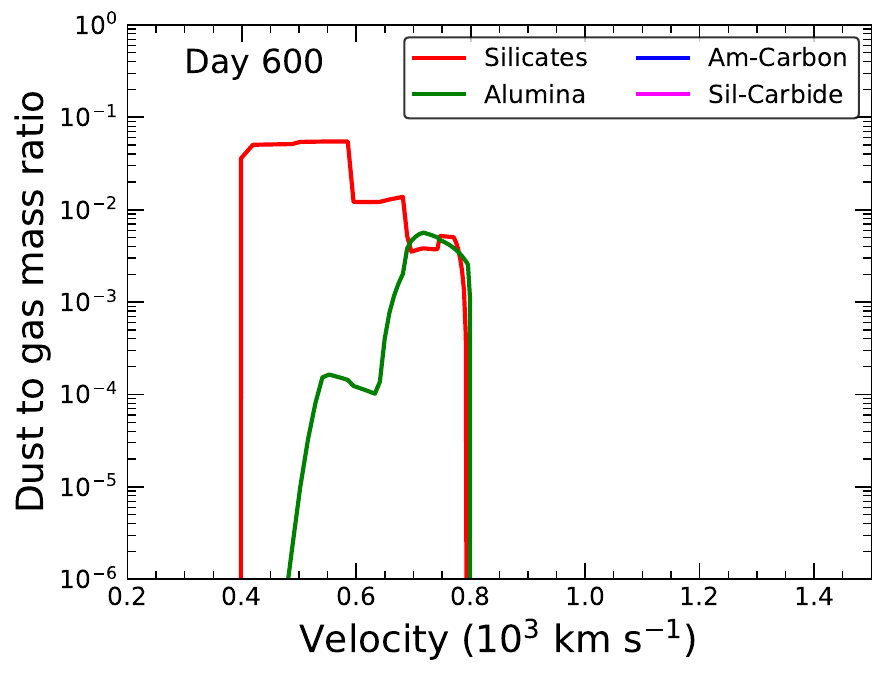}
\includegraphics[width=3.5in]{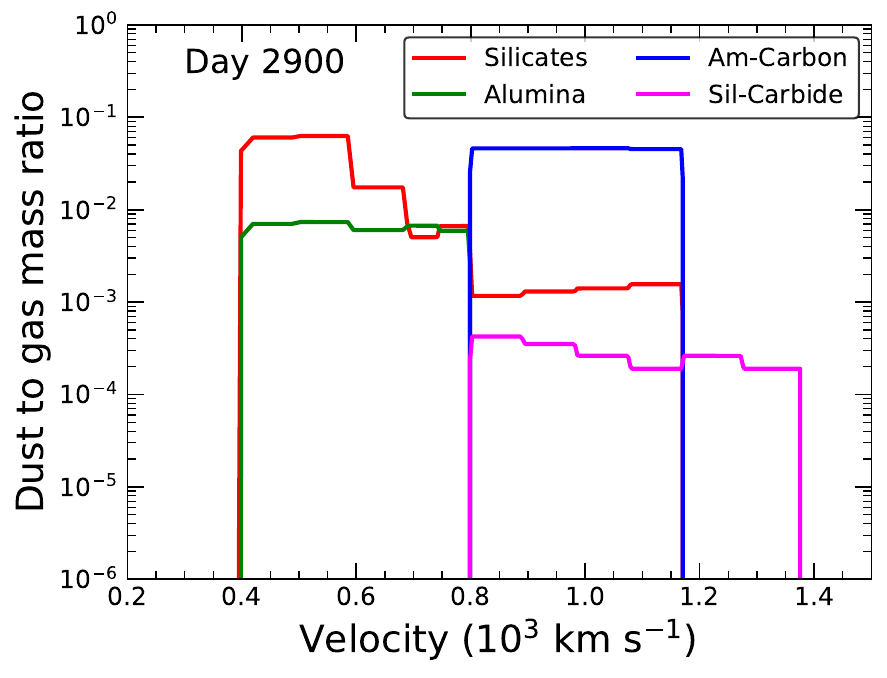}
\caption{\label{fig_model}\footnotesize{\textit{Top}: The composition and evolution of dust in SN~2005af, derived from the dust-formation model, is presented. O-rich dust, namely silicates and alumina and C-rich dust, namely amorphous carbon and silicon carbide, are considered. The model suggests steady growth of O-rich dust between days 400 and 800 post-explosion. Amorphous carbon dust forms at a rapid rate after 1100 days, and it becomes the most abundant dust component in the ejecta. Based on this study, we suggest that in SN~2005af, the ejecta are predominantly C-rich in dust. \textit{Bottom}: The suggested distribution of dust within the ejecta of SN~2005af is shown for day 600, when dust formation has just begun, and for day 2900, when the formation of dust is already saturated. It is evident that O-rich dust is likely to form in the inner regions, while C-rich dust forms in the outer parts of the He core. }} 
\end{figure*}

\begin{table}
\centering
\caption{Progenitor and explosion properties of SN~2005af derived from light-curve modeling. 
\label{tab:table_lightcurve}}
\begin{tabular}{cccc}
\hline
\hline
Progenitor & Ejecta & E$_{\rm exp}$ & \Ni \\
\Ms & \Ms & erg s$^{-1}$ & \Ms \\
\hline
10.0  & 8.3 & 0.39 $\times$ 10$^{51}$ & 0.050 \\
\hline
\end{tabular}
\end{table}

\begin{figure}
\vspace*{0.3cm}
\includegraphics[width=3.5in]{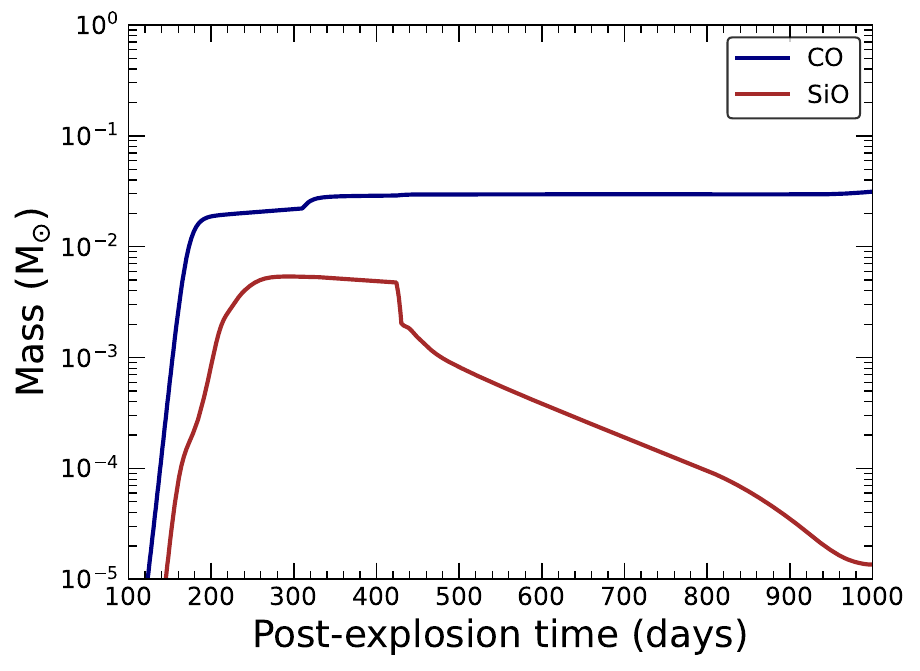}

\caption{\label{fig_molecules}\footnotesize{The model predictions for the mass of molecules CO and SiO as a function of post-explosion time in the ejecta of SN~2005af. SiO molecules are considered a tracer and precursor of silicate dust formation \citep{sar13}, while CO molecules remain as the most abundant molecule in the gas. }} 
\end{figure}

The top panel of Figure \ref{fig_model} shows the evolution of dust in the ejecta of SN~2005af. The top panel displays the distribution of dust in velocity space at days 600 and 2900. We find that the O-rich dust, such as silicates, forms rapidly (as early as day 400), and the mass of silicates at day 600 is already $4 \times 10^{-3}$ \Ms. Silicates and alumina form in the inner O/Si-rich core, which in this case is confined only within an expansion velocity of 1200 km s$^{-1}$. Carbon and silicon carbide forms much later, after 1100 days post-explosion. Rapid condensation of carbon dust takes place in the outer He/C-rich layers (velocities between 800 and 1400 km s$^{-1}$) at that time. We find that, overall, C-rich dust will dominate the dust composition of SN~2005af, with 0.02 \Ms\ of amorphous carbon dust formed in the ejecta. This is expected to be a characteristic of low-mass progenitors, given their relatively lower mass of the O-rich layer compared to the  He-rich layer \citep{sar13, sukhbold_2016}. For the model, the total dust mass is about 0.03 \Ms\ formed in the metal-rich He core of the ejecta.  Molecules such as CO and SiO are abundantly formed in the ejecta, as shown in Figure \ref{fig_molecules}. While CO is the most abundant molecule, SiO depletes rapidly after day 400, leading to the formation of Silicate dust. 

These theoretical predictions are used in the following sections as a reference when analyzing the IR data for dust composition and location.

\begin{figure*}
\vspace*{0.3cm}
\centering
\includegraphics[width=6.5in]{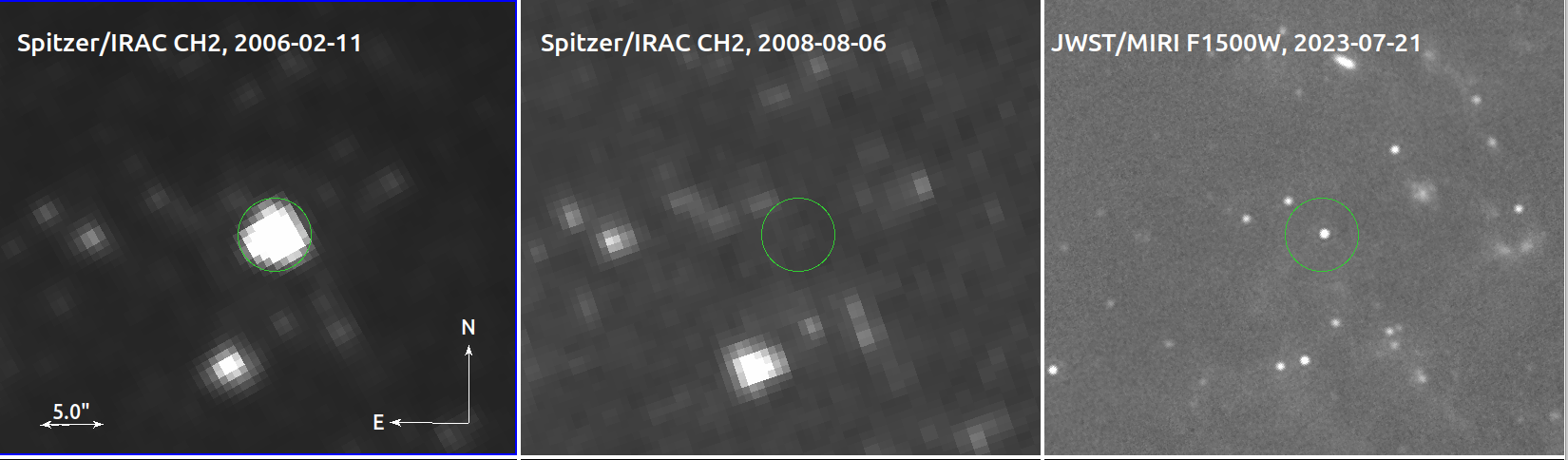}
\caption{\label{fig_Spitzer_image}\footnotesize{\textit{Left and Middle}: Spitzer IRAC images of SN~2005af taken on February 2006 (458~days post-explosion) and August 2008 (1363~days post-explosion) \citep{Szalai13} \textit{Right}: JWST/MIRI image of SN~2005af taken in July 2023 (6826~days post-explosion) in filter band F1500W. }}
\end{figure*}

\begin{figure}
\vspace*{0.3cm}
\centering
\includegraphics[width=3.5in]{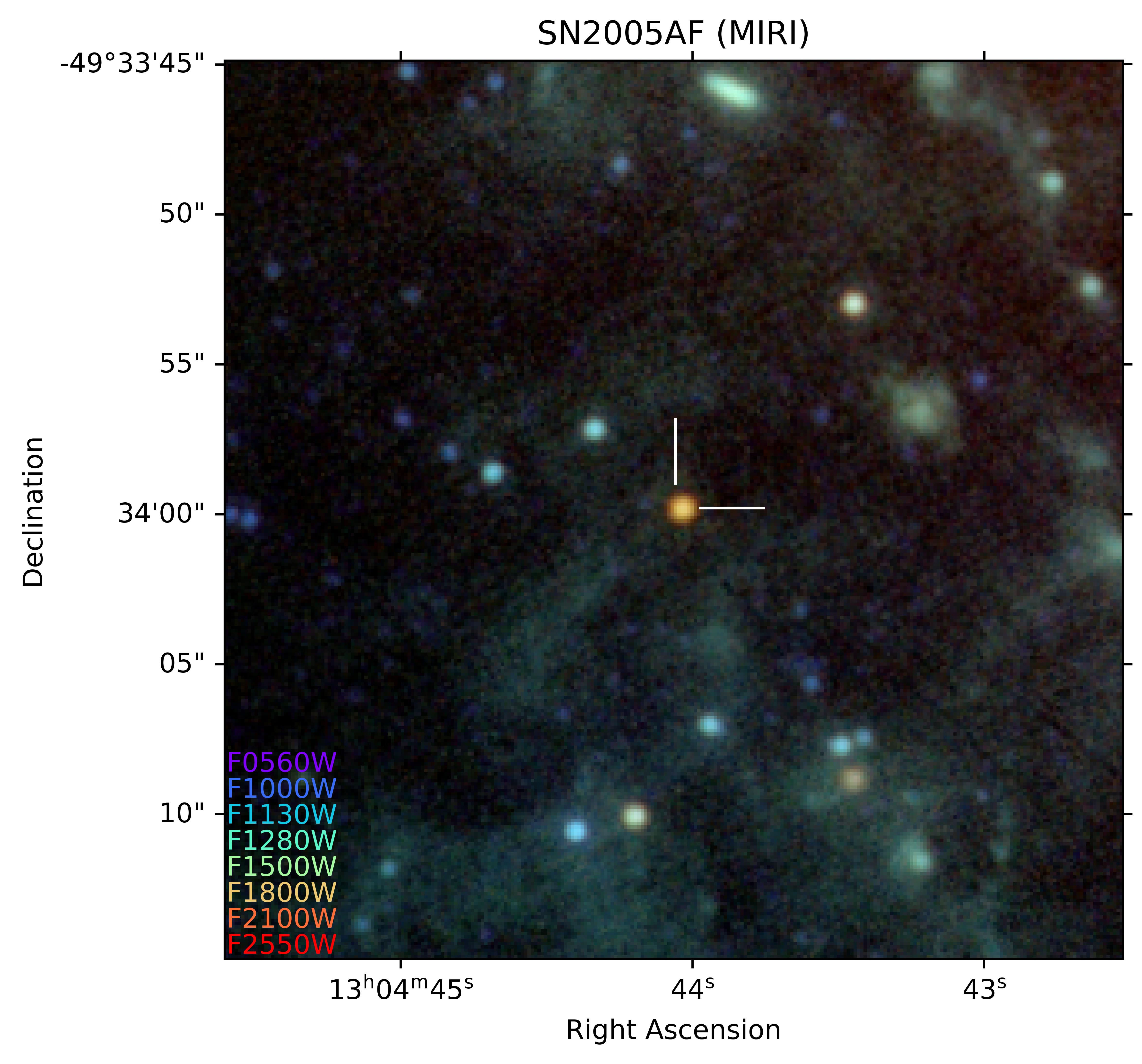}

\caption{\label{fig_JWST_MIRI}\footnotesize{JWST/MIRI composite image of SN~2005af, taken on 2023-07-21, which is 6826 days post-explosion. The white tick marks show the position
of the supernova.}} 
\end{figure}

\section{Optical data}
\label{sec_optical}

In this study, we used the optical photometric data of SN 2005af (see Figure \ref{fig_lightcurve}) in the \textit{B} (0.435\mic), \textit{g} (0.477~\mic), \textit{V} (0.538~\mic), \textit{r} (0.622~\mic), and \textit{i} (0.761~\mic) bands, corresponding to epochs between 92 and 163 days post-explosion \citep{anderson_2024}, obtained as part of CSP-I \citep{humay_2006}. We also present a single unpublished late-time optical spectrum of SN~2005af (Figure \ref{fig_lightcurve}, right-panel) obtained with the Keck Low Resolution Imaging Spectrometer (LRIS; \citealp{Oke_1995}) on 2022 March 4 (at epoch 6322 days).
The spectrum was acquired with the slit oriented at or near the parallactic angle to minimise slit losses caused by atmospheric dispersion \citep{Filippenko_1982}. 
The LRIS observation utilized the $1\arcsec$-wide slit, 600/4000 grism, and 400/8500 grating to produce a similar spectral resolving power ($R \approx 700$--1200) in the red and blue channels.

\section{\textit{Spitzer} Observations}
\label{sec_Spitzer}

\subsection{Photometry}
SN~2005af was imaged several times with the {\it Spitzer} Infrared Array Camera (IRAC) and Multiband Imaging Photometer for {\it Spitzer} (MIPS) between 2005 and 2019 (Figure \ref{fig_Spitzer_image}). As noted in Section \ref{sec:intro}, we have updated the explosion epoch to be MJD 53320.8 based on \cite{gutierrez_2017}, hence the epochs mentioned previously by \cite{Szalai13} and \cite{kotak_2006} are offset (increased) by 58 days.

For SN~2005af, the early evolution, \citet{kotak_2006} and \citet{kot08} presented mid-IR observations. 
\citet{kotak_2006} reported the first identification of the SiO molecule in a Type IIP SN, based on observations from day 272. \citet{Szalai13} presents the full {\it Spitzer} dataset of SN~2005af up to $\sim$ 998 post-explosion. That study, however, is based on the analysis of the Post Basic Calibrated Data (PBCD) -- applying simple aperture photometry with a 5-12-20 configuration in native 1.2\arcsec IRAC pixels --, which results in a $\gtrsim 10\%$ photometric uncertainty.
Thus, for our current study, we repeated the whole analysis: we downloaded the Basic Calibrated Data (BCD) of the SN from the {\it Spitzer} Heritage Archive (SHA) and thoroughly reanalyzed all the early-time archival {\it Spitzer} IRAC data
(but we did not use the MIPS images because of their  spatial and spectral limitations).
During the photometric reduction, we performed overlap correction and mosaicing of the frames with {\tt MOPEX} \citep{makovoz05}. We invoked the task {\tt APEX} User List Multiframe,
with the absolute SN position as input, to extract the flux via
fitting of the point-response function (PRF) for each epoch. Note that the formal uncertainties of the fluxes (based on the signal-to-noise ratio) are implausibly low.
Thus, to obtain more realistic error measurements, we used the uncertainties derived from an alternative aperture-photometry analysis (described by \citealt{Fox11}) conducted on both the source and the background.
With this method, we are able to sample the background of just the SN  and eliminate the effect of bright sources in its vicinity.
The results of the {\it Spitzer}/IRAC photometry of SN~2005af are presented in Table \ref{tab:Spitzer_phot}. As can be seen, the new photometry is in good agreement with the values published in \citet{Szalai13} at the first two epochs; however, there are larger differences at later times (probably because of the too large aperture size used in the previous study), which confirms our strategy on revising the previously published photometry.

\begin{table*}[]
\centering
\caption{The results of the {\it Spitzer}/IRAC photometry of SN~2005af. Numbers marked with italic show the photometric values previously published in \cite{Szalai13}.}
\label{tab:Spitzer_phot}
\begin{tabular}{lccccc}
\hline
\hline
\centering
MJD & Epoch & F$_{3.6}$ & F$_{4.5}$ & F$_{5.8}$ & F$_{8.0}$ \\
(days) & (days) & ($\mu$Jy) & ($\mu$Jy) & ($\mu$Jy) & ($\mu$Jy) \\
\hline
53573.3	& 252.5 & 3772$\pm$100 & 17940$\pm$216 & 9284$\pm$147 & 6968$\pm$112 \\
 & & {\it (3849$\pm$104)} & {\it (17984$\pm$211)} & {\it (9014$\pm$151)} & {\it (7283$\pm$138)} \\
53778.9 & 458.1 & 858$\pm$44 & 3210$\pm$88 & 2139$\pm$59 & 4448$\pm$80 \\
 & & {\it (956$\pm$52)} & {\it (3348$\pm$91)} & {\it (2192$\pm$75)} & {\it (4870$\pm$115)} \\
53955.1 & 634.3 & 136$\pm$13 & 407$\pm$27 & 848$\pm$35 & 1769$\pm$38 \\
 & & {\it (254$\pm$28)} & {\it (492$\pm$35)} & {\it (972$\pm$52)} & {\it (2193$\pm$78)} \\
54151.12 & 830.3 & 13$\pm$3 & 25$\pm$4 & 91$\pm$5 & 256$\pm$8 \\
 & & {\it (135$\pm$22)} & {\it (112$\pm$18)} & {\it (236$\pm$29)} & {\it (774$\pm$55)} \\
\hline
\end{tabular}
\end{table*}

\subsection{Spectroscopy}
SN~2005af has been also observed with the {\it Spitzer} InfraRed Spectrograph (IRS) at four epochs \citep{kotak_2006, kot08} during its early evolution (at days 125, 272, 629, and 864 after explosion).
We downloaded the four available IRS spectra of SN~2005af from the Combined Atlas of Sources with Spitzer IRS Spectra
(CASSIS)\footnote{\href{https://cassis.sirtf.com/}{https://cassis.sirtf.com/}} database \citep{Lebouteiller11}.
We used the best flux-calibrated spectra, respectively, following the guidelines provided by the CASSIS recommendations.

\section{{\sl JWST\/}/MIRI Imaging}\label{section_MIRI}

As part of the Cycle 1 General Observers (GO) 2666 program\footnote{\href{https://www.stsci.edu/jwst/science-execution/program-information?id=2666}{https://www.stsci.edu/jwst/science-execution/program-information?id=2666}} (PI O.~D. Fox), we obtained images of SNe~2005af with the {\it JWST} Mid-Infrared Instrument (MIRI) on 2023-07-21 ( 6826 days post-explosion).
The observations were acquired in the F560W, F1000W, F1130W, F1280W, F1500W, F1800W, F2100W, and F2550W filter bands, using the FASTR1 readout pattern in the FULL array mode and a 4-point extended source dither pattern (Figure \ref{fig_JWST_MIRI}). 
A description of our detailed calibration process of the {\it JWST}/MIRI images was recently published by \citet{shahbandeh_2023}. We use the JWST HST Alignment Tool \citep[JHAT;][]{Rest_2023} to align {\it JWST} and {\it HST} images of the fields of the host galaxy with each other.

To measure the fluxes of SN~2005af on {\it JWST}/MIRI images, we followed the method described in detail by \citet{shahbandeh_2023}. 
We performed point-spread-function (PSF) photometry on background-subtracted level-two data products using \texttt{WebbPSF} \citep{Perrin_2014} implemented in the \texttt{space-phot} package\footnote{https://zenodo.org/records/12100100} \citep{pierel_2024_12100100}.
In order to calibrate the flux, we applied flux offsets by measuring the PSF of all the stars in the field and comparing them to the corresponding catalogues created by the pipeline. 
The fluxes of all four dithers of each filter were then averaged. 
The final results of {\it JWST}/MIRI photometry of SN~2005af are presented in Table~\ref{tab:JWST_phot}.


\begin{table}
\centering
\caption{JWST/MIRI Photometry of SN~2005af \label{tab:JWST_phot}}
\setlength{\tabcolsep}{3pt}
\begin{tabular}{ l c c}
\tableline
\tableline
\centering
Filters & AB Mag & F$_\nu$ ($\mu$Jy) \\
\tableline
F560W & $<$ 22.03 & $<$ 5.6 \\
F1000W & 22.01 $\pm$ 0.10 & 5.7 $\pm$ 0.6 \\
F1130W & 20.80 $\pm$ 0.10 & 17.3 $\pm$ 1.5 \\
F1280W & 19.80 $\pm$ 0.03 & 43.6 $\pm$ 1.1 \\
F1500W & 18.89 $\pm$ 0.01 & 100.7 $\pm$ 1.4 \\
F1800W & 18.32 $\pm$ 0.02 & 171.0 $\pm$ 2.9 \\
F2100W & 17.95 $\pm$ 0.02 & 240.2 $\pm$ 5.2 \\
F2550W & 16.93 $\pm$ 0.03 & 613.7 $\pm$ 17.8 \\
\tableline
\end{tabular}
\tablecomments{All observations were taken on 2023-07-21, that is 6826 days post-explosion.}
\end{table}



\section{Evolution of dust}
\label{sec_Spitzer_model}

To describe the dust formation history of the SN we fit its mid-IR emission. The distance of the SN was taken as 3.9~Mpc \citep{jacques_2005, filippenko_2005, kotak_2006}. 
The nature and evolution of dust in SN~2005af from 2~months post-explosion, until 18.7~yr is estimated using $Spitzer$ and {\it JWST} data summarized above. We applied both blackbody and analytical dust models. The parameters are listed in Table \ref{table_dustfit}. We have used silicates and amorphous carbon dust as the two main dust types, whose optical properties are given by \cite{dra07} and \cite{zub04}. Here we have derived the amount of pre-existing dust in the CSM, which was formed in the pre-explosion winds, in addition to the amount of newly formed dust in the SN ejecta, post-explosion. For simplicity, we have chosen all grains to be of spherical shape and 0.1 $\mu$m radius, as also previously assumed by \cite{sarangi_2022b} and \cite{shahbandeh_2023}. To resolve the degeneracies while fitting the IR data, such as when multiple scenarios can produce a reasonably good fit, we have relied on the physical models (Section \ref{sec_model}) for determining the most likely scenario.

\subsection{Dust Emission and Geometry}

We estimated the mass and temperature of dust based on the assumptions that (a) the SN ejecta and the CSM have spherically symmetric geometry, (b) dust at a given location has identical temperature, (c) and the dust grains are of identical size $a$ of 0.1 \mic. As argued in previous studies \citep{fox_2010, sarangi_2022b, shahbandeh_2023}, in the mid-IR regimes, grain size distributions do not have a significant impact on the fluxes. 

At a given time, the flux produced by dust of mass $m_d$ and temperature $T_d$ is given by 
\begin{equation}
\label{eq_fluxfit}
F_{\lambda}(\lambda, T_d) =  \frac{4 \ m_d \ k(\lambda, a) \ \pi B_{\lambda} (\lambda, T_d) P_{\rm esc}(\lambda)}{4 \pi D^2}\, ,
\end{equation}
where $D$ is the distance of the SN, $B$ is the Planck function, $k(\lambda,a)$ is the mass absorption coefficient, and $P_{\rm esc}(\lambda)$ is the escape probability of a photon from the dust cloud. We have used the absorption coefficients $k(\lambda)$ for silicates and amorphous carbon dust from \cite{dra07} and \cite{zub04}, respectively, similar to our recent studies \citep{szanna_2024, shahbandeh_2023, sarangi_2022b}. The escape probability of a photon of wavelength $\lambda$ from a dusty sphere of uniform temperature \citep{dwe15, inoue_2020} is given by

\begin{equation}
P_{esc}(\lambda) = \frac{3}{4\tau} \Big[1 - \frac{1}{2\tau^2} + \big (\frac{1}{\tau} + \frac{1}{2\tau^2} \big)e^{-2\tau} \Big]\, .
\end{equation}

The optical depth $\tau(\lambda)$ of a sphere of radius $R$ with uniform dust density is given by

\begin{equation}
\label{eq_tausph}
\tau(\lambda) = \ \frac{3 m_d k(\lambda)}{4 \pi R^2}\, .  
\end{equation}
\noindent
If the dust is distributed in a spherical shell of inner and outer radii $R_1$ and $R_2$ (respectively), the optical depth is given by

\begin{equation}
\label{eq_taushell}
\tau(\lambda) = \frac{3 m_d k(\lambda)}{4 \pi (R_2^2 + R_1 R_2 + R_1^2)}\, . 
\end{equation}
\noindent
When the mass of dust is large, the respective optical depth is large as well. In that case, the escape probability at any given wavelength is small, since the IR radiation from the dust is self-absorbed within the cross-section of the dusty sphere. In the other extreme scenario, where the mass of dust is small, representing an optically thin medium, the escape probability is close to 1. 

We have used Equation \ref{eq_fluxfit} to fit the obtained IR fluxes of SN~2005af, as described below. The \texttt{python scipy} library \texttt{curve\_fit} was used for fitting and error analysis. 

\begin{figure*}
\vspace*{0.3cm}
\centering
\includegraphics[width=3.5in]{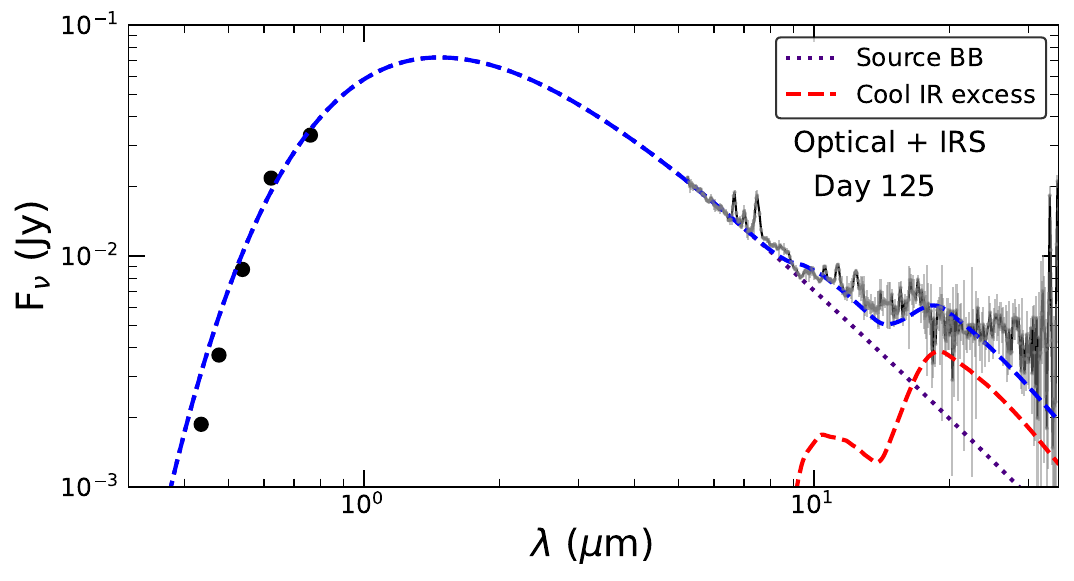} 
\includegraphics[width=3.5in]{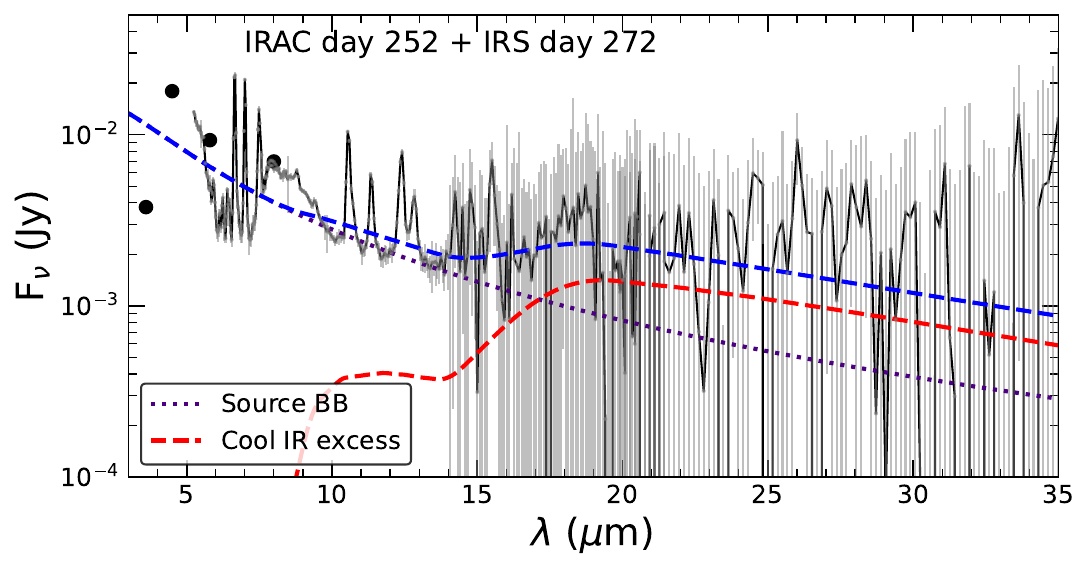} 
\includegraphics[width=3.5in]{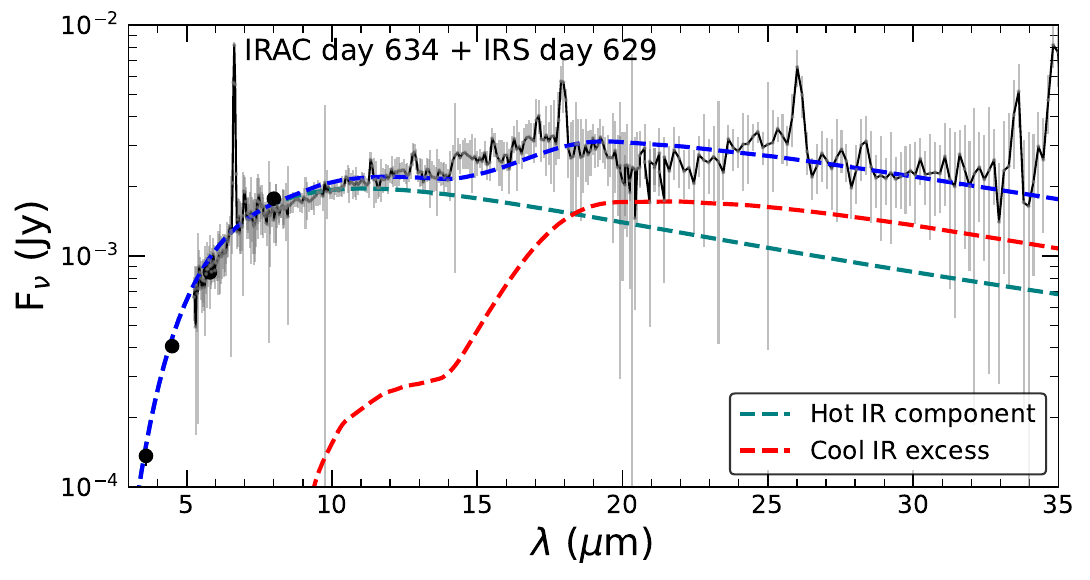} 
\includegraphics[width=3.5in]{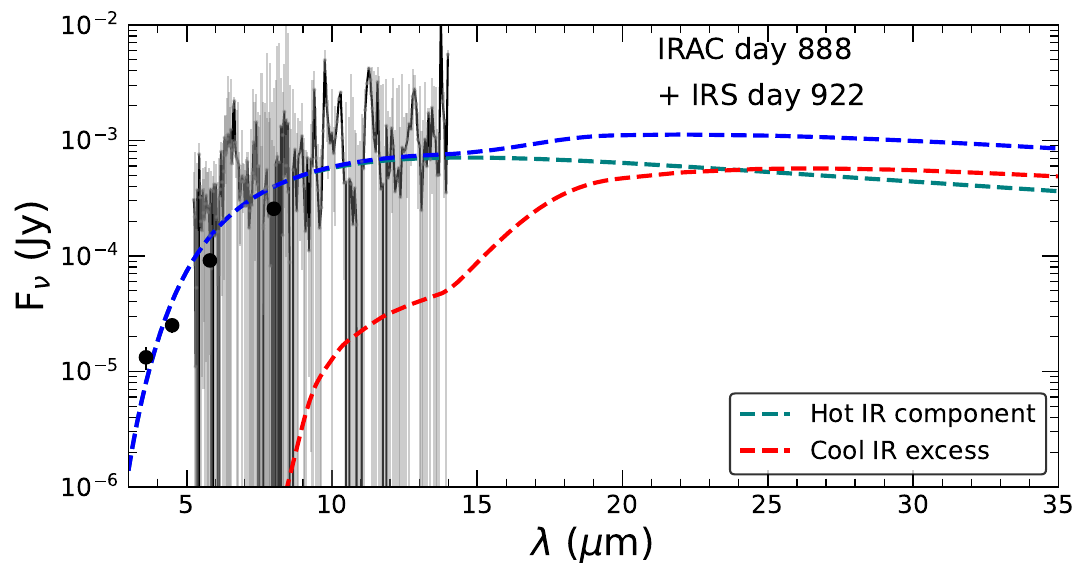}
\caption{\label{fig_spitzer}\footnotesize{The best-fit cases for four epochs of observation by {\it Spitzer} through IRAC imaging (solid black circles) and IRS spectroscopy. For day 125, we combined the optical data from CSP-I with the {\it Spitzer} spectrum. The epochs of day 125 and day 272 (+ IRAC day 252) were fit using a blackbody (in dotted, purple line) for the optical component, along with a cool IR component (dashed, red line) from dust (sum of the two components given in blue). For epoch 629 (+ IRAC day 634) and epoch 806 (+ IRAC day 830), we assumed a combination of hot (in green) and cool IR (red) components. We suggest that the cool IR component for all the cases originates from some pre-existing dust in the circumstellar matter, which was formed in the winds of the progenitor before the explosion. The hot IR component is attributed to emission from newly formed dust in the ejecta. See Table \ref{table_dustfit} for the fitting parameters and Section \ref{sec_Spitzer_model} for a detailed description of the scenario. }} 
\end{figure*}

\begin{figure}
\vspace*{0.3cm}
\centering
\includegraphics[width=3.5in]{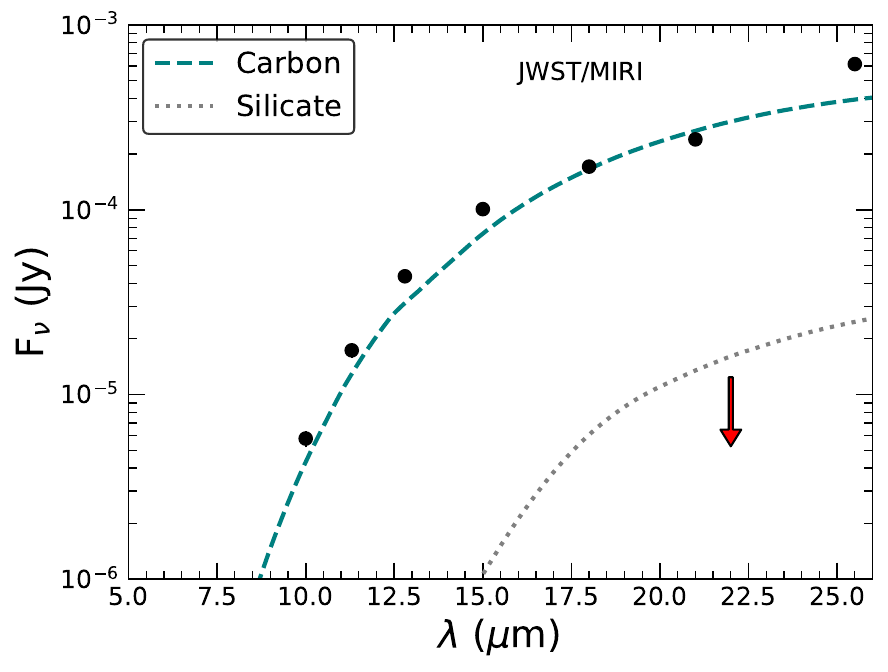}
\caption{\label{fig_JWST}\footnotesize{The best-fit model is presented for mid-IR fluxes (data given in Table \ref{tab:JWST_phot}) of SN~2005af obtained through {\it JWST}/MIRI, 6826 days post-explosion. We find that, at this epoch, dust is predominantly C-rich in composition. See Section \ref{sec_JWST_model} and Table \ref{table_dustfit} for the description. }} 
\end{figure}

\subsection{Day 125}


At day 125, the spectrum is largely dominated by the optical emission from the SN photosphere. We combined the optical photometric data obtained from CSP-I (Figure \ref{fig_lightcurve}) at epoch 125 post-explosion, along with the \textit{Spitzer} spectrum at the same epoch. The optical part of the data was fit using a blackbody spectrum. The blackbody component is of temperature 3459~K and radius 2.1$\times$10$^{15}$~cm.
In addition, we find the presence of a significant excess in the near-IR and mid-IR, dominating over the optical blackbody (Figure \ref{fig_spitzer}, Table \ref{table_dustfit}). Such early epochs are not suitable for dust condensation in the hot ejecta, as seen in Section \ref{sec_model} \citep{sarangi2018book}. \cite{pereyra_2006} suggests a possibility for the presence of a dusty CSM in SN~2005af. We consider this emission to be from the dust in the CSM, which was formed in the winds prior to the explosion. In Type II-P SNe, the CSM is formed by the winds blown from the H-rich outer envelope of the massive star, where the C/O ratio is smaller than 1 \citep{sukhbold_2016}. O-rich dust such as silicates is expected in such environments \citep{lebouteiller_2012, clark_2013}. We find the mass of silicate to be about 2.9 $\times$ 10$^{-3}$~\Ms, at 179~K. At this early epoch, such low dust temperatures require the dust to be present at large distances from the SN. Our calculation suggests the dust to be around 2-4$\times$10$^{17}$~cm from the photosphere, to match this temperature. In comparison, if we assume a forward shock velocity of 5000 km s$^{-1}$, the SN shock will be only at 5$\times$10$^{15}$~cm at this epoch. Therefore, the dust is very likely to be pre-existing in the CSM.  Ejecta dust, if present within the photosphere at these early epochs, is supposed to be much hotter. 

\subsection{Day 272}

Day 272 shows a trend similar to that of day 125, where the combined \textit{Spitzer}/IRAC fluxes and the IRS spectrum show the presence of an IR-excess (Figure \ref{fig_spitzer}, Table \ref{table_dustfit}). In sync with the findings of day 125, we report that this is a pre-existing component of dust in the CSM, which gives rise to the IR-excess. We do not expect significant dust condensation at such early times. However, due to the lack of optical data and the poor quality of the obtained \textit{Spitzer} spectrum, our fit is not very reliable. The temperature of the optical blackbody was obtained to be 2400~K, and a radius of 1.6$\times$10$^{15}$~cm.  We find the mass of silicate dust to be $\sim 2.0\times 10^{-3}$~\Ms, at 157~K. Like in the case of day 125, this dust temperature corresponds to a distance of 2$\times$10$^{17}$~cm, while the SN shock is expected be only around 10$^{16}$~cm at this epoch. On day 272, molecular emission from SiO was detected in its fundamental band in the range 7.9--9~\mic, which is a precursor to dust condensation in the ejecta \citep{kotak_2006, sar13}. Note that according to our models, the O-rich dust forms rapidly around day 400.  A part of the mid-IR spectrum at this epoch is therefore expected to have some contribution from molecular lines as well. 

Importantly, \citet{kotak_2006} analyzed the same \textit{Spitzer} data but did not report the presence of dust emission in the spectra at either day 125 (day 61 in \citealt{kotak_2006}) or day 272 (day 214 in \citealt{kotak_2006}). We agree that atomic and molecular line emission from CO and SiO in the ejecta may dominate the spectra until 9.5~\mic, as noted by \citet{kotak_2006}. We did not include molecular emission in our fit. However, we report evidence for a dust component in the extended CSM gas, which contributes to an IR excess in the spectra at both epochs. This excess is most significant at longer wavelengths beyond 10~\mic\ and does not conflict with the presence of molecules. In fact, our dust-evolution scenario supports the appearance of SiO molecules at day 272, which subsequently lead to the formation of silicate dust in the ejecta at later epochs. The mid-IR data from day 272 is noisy. But since we find compelling evidence of an IR excess at day 125, arguably originating from pre-existing dust, we suggest that the component should also be present at day 272.

\begin{table*}

\centering
\caption{Best-fit parameters and percentage errors for all analyzed IR observations of SN~2005af spanning over two decades, compared to model results. \label{table_dustfit}}
\begin{tabular}{l l c c c | c }
\hline \hline
Parameter & Unit &  Optical BB & Pre-existing CSM  & Ejecta dust & Ejecta model\\
 & &  (error\%) & dust (error\%) & (error\%) & \\
\hline
\multicolumn{6}{c}{Day 125 \textit{Spitzer}/IRS} \\
\hline
BB radius & cm & 2.1$\times$10$^{15}$ (0.1) & --& --& -- \\
Dust type & -- & -- & Silicate &-- & -- \\
Temperature & K & 3459 (0.4) & 179 (8) & -- & -- \\
Dust mass & \Ms & -- &2.9$\times$10$^{-3}$ (38) & -- & -- \\
Dust velocity & km s$^{-1}$ & -- & -- & -- & -- \\

\hline
\multicolumn{6}{c}{Day 272 \textit{Spitzer}/IRS + IRAC} \\

\hline

BB radius & cm & 1.6$\times$10$^{15}$ (0.1) & -- & -- & -- \\
Dust type & -- & -- & Silicate &-- & -- \\
Temperature & K & 2400 (6.9) & 157 (51) & -- & -- \\
Dust mass & \Ms & -- & 2.0$\times$10$^{-3}$ (246) & -- & -- \\
Dust velocity & km s$^{-1}$ & -- & -- & -- & -- \\
\hline
\multicolumn{6}{c}{Day 629 \textit{Spitzer}/IRS + IRAC} \\

\hline

Dust type & -- & -- & Silicate & Silicate & Silicate \\
Temperature & K & -- & 129 (5) & 459 (6) & -- \\
Dust mass & \Ms & -- &6.5$\times$10$^{-3}$ (23) & 3.0$\times$10$^{-3}$ (min.) & 4.2$\times$10$^{-3}$ \\
Dust velocity & km s$^{-1}$ & -- & -- & 1290 (10) & 810 \\
\hline

\multicolumn{6}{c}{Day 830 \textit{Spitzer}/IRAC + Day 864 IRS} \\

\hline

Dust type & -- & -- & Silicate & Silicate & Silicate \\
Temperature & K & -- & 106 (95) & 349 (31) & -- \\
Dust mass & \Ms & -- &6.0$\times$10$^{-3}$ & 2.2$\times$10$^{-3}$ (min.) & 4.8$\times$10$^{-3}$ \\
Dust velocity & km s$^{-1}$ & -- & -- & 800 (79) & 814 \\
\hline

\multicolumn{6}{c}{Day  6826 (18.7 years) {\it JWST}/MIRI} \\

\hline

Dust type & -- & -- & -- & Carbon, Silicate & Carbon, Silicate \\
Temperature & K & -- & -- & 108 (76), 80 (max.) & -- \\
Dust mass & \Ms & -- & -- & 0.0143 (64), 1.8$\times$10$^{-3}$ & 0.021, 4.7$\times$10$^{-3}$ \\
Dust velocity & km s$^{-1}$ & -- & -- & 1250 (1100), 706 & 1190, 805 \\
\hline

\end{tabular}
\end{table*}

\subsection{Day 629}

On day 629, we find the presence of two distinct IR components, hot and cold (Figure \ref{fig_spitzer}, Table \ref{table_dustfit}). The cold silicate component at 129~K with mass of $\sim$ 6.0 $\times$ 10$^{-3}$~\Ms is aligned in temperature with the pre-existing dust that produces the IR excess at days 125 and 272. Even though the CSM dust mass should not increase, a larger mass estimated from the spectra may be attributed to the increase in the echo due to the effect of the light-travel time (\citealt{dwek_1985, dwek_2021}). 

Most core-collapse SNe are known to form dust by day 629 \citep{luc89, sarangi2018book}, and the chemical model of SN~2005af in Section \ref{sec_model} predicts the formation of silicate dust of $\sim 4.2 \times 10^{-3}$~\Ms\ in the ejecta. Moreover, the detection of SiO molecules \citep{kotak_2006, crowther_2007} also suggests the formation of silicate dust in this SN. However, the spectrum does not show any notable peak around 9.7~\mic, which is characteristic of silicate dust. We find that, if at least $3.0 \times 10^{-3}$~\Ms\ of silicate dust is formed in the O-core of the ejecta expanding at $\sim 1290$~km~s$^{-1}$ (assuming homologous expansion), the emission from that hot newly formed dust ($\sim 459$~K) can fit the observed spectrum. A blackbody radius for this spectrum results in a velocity of $\sim 1200$~km~s$^{-1}$ at day 629. Hence, an emission from optically thick silicate with a similar velocity can well explain the origin of the hot component of the IR spectrum. Owing to the dust being optically thick (Eq. \ref{eq_tausph}), it suppresses the 9.7 and 18~\mic\ silicate features in the emission spectrum \citep{dwe15, sarangi_2022b}. The dust mass is quite in agreement with the theoretical model, where $4.2\times 10^{-3}$~\Ms\  of dust is found to have formed inside a sphere of $\sim 810$ km s$^{-1}$. Technically, there is no threshold mass that defines when the dusty sphere becomes optically thick, since optical depths are wavelength dependent. We find in this case that any dust mass larger $3.0 \times 10^{-3}$~\Ms\ does not change the total IR luminosity, and hence we can consider it as a reasonable lower limit for the dust mass in the ejecta. 

\subsection{Days 830 and 864} 

The IRAC fluxes on day 830 and the IRS spectrum on day 864 are combined in a single analysis. However, the data are subject to large uncertainties. The silicate dust in the ejecta is taken as responsible for the hot IR emission. For the best scenario, we find silicates in the ejecta have a mass lower limit of $2.2 \times 10^{-3}$~\Ms\ and a temperature of 349~K. The pre-existing dust mass remains unaltered. Given the large uncertainties in the observed fluxes, we suggest the reader to only consider the trend and not stress on the absolute values from the fit at this epoch. 




\subsection{Day  6826 - Year 18.7 (JWST/MIRI)}
\label{sec_JWST_model}

SN~2005af was observed by {\it JWST} in 2023,  about 19~yr after the explosion and $\sim 16$~yr from when it was last observed with $Spitzer$. The fluxes in the MIRI bands (10.0, 11.3, 12.8, 15.0, 18.0, 21.0, and 25.5 \mic) do not show any noticeable spectral features (Figure \ref{fig_JWST}). We find that the data fit well with an amorphous carbon component, while silicate dust is not a good candidate given its strong features at 9.7 and 18~\mic. For clarity of understanding, in the first few years after the explosion, the expanding SN ejecta remain very compact in radius, hence, a small mass of dust makes it optically thick at mid-IR wavelengths. It is difficult to distinguish between carbon and silicate dust at that point, since the silicate features are suppressed due to large opacities. However, at later times, such as 19~yr, the ejecta expand to a large radius, and hence the mid-IR optical depths are reduced considerably. At these epochs, it is possible to determine the likely dust composition. 
We find a mass of carbon dust to be 0.014~\Ms\ at a temperature of 108~K (Figure \ref{fig_JWST}, Table \ref{table_dustfit}). The IR luminosity at this epoch is $\sim 10^{38}$~erg~s$^{-1}$. The velocity of the dusty sphere is found to be 1250~km~s$^{-1}$. However, owing to the featureless nature of the fluxes, the velocity is not well constrained, incurring large uncertainties. The model presented in Section \ref{sec_model} predicts the mass of carbon dust to be 0.02~\Ms, where amorphous carbon is the most abundant dust type in SN~2005af. The velocity of the dusty region (1190~km~s$^{-1}$) is also in agreement with the fit to the observed data. 

At the \textit{Spitzer} epochs, we considered the presence of silicate dust in the ejecta. When analyzing the {\it JWST} fluxes, we find that carbon and silicates may coexist in the ejecta; however, the silicate must be cooler than the amorphous carbon dust. The best-fit temperature for carbon is 108~K, and the maximum possible temperature for silicates is $\sim 80$~K. In Figure~\ref{fig_JWST}, we show the maximum silicate contribution with a downward arrow. The likely scenario where carbon and silicate dust are at different temperatures cannot be addressed based on the data alone. With regard to the stellar structure \citep{rau02, woosley1989, sukhbold_2016} and the dust-formation zones, as presented in Section \ref{sec_model} \citep{sar13, sarangi_2022b}, it has been predicted that carbon dust forms in the outer He-rich layers, while O-rich silicate dust forms essentially in the O-core. Analyzing the {\it JWST} data from SN~2004et and SN~2017eaw, it has been concluded that the temperature of the ejecta dust after a few years is controlled by heating from an external forward shock \citep{shahbandeh_2023}. We find that 0.014~\Ms\ of carbon dust in the outer shell at velocities of 800--1200~km~s$^{-1}$ has optical depths $\tau$(0.1 \mic), $\tau$(0.3 \mic), and $\tau$(0.6 \mic) of 2.1, 2.6, and 2.9, respectively (Eq. \ref{eq_taushell}). Therefore, if silicate dust is internal to carbon dust and is heated by an external shock (see Figure \ref{fig_results}), we can justify the difference in temperature. In the future, we will construct a complete radiative-transfer model to calculate the variation of dust temperatures between zones. 

We note that there is an excess at the longest MIRI filter of 25.5~\mic, over our fit with 0.014~\Ms\ of carbon dust at 106~K, as visible in Figure \ref{fig_JWST}. This indicates that there might be larger masses of cold dust that are not detected by {\it JWST}. SN~1987A was found to host about 0.5 \Ms\ of dust at a temperature of 17-23~K \citep{matsuura_2015}. The dust mass we detected by {\it JWST}, by analyzing the mid-IR emission, is a lower limit of the dust present in SN~2005af.

\subsection{Summary of the Dust Evolution}
\label{sec_summary_dust}

 In Figure \ref{fig_results}, we present a simple schematic scenario of the location, type, and amount of dust that is present in SN~2005af after 20~yr post-explosion. This is the first study where the dust formed in the CSM prior to the explosion and newly formed dust in the ejecta are differentiated, combining modeling and observations. We find amorphous carbon to be the dominant dust species in the ejecta, constituting almost 80\% of the total dust. This is justified if SN~2005af was indeed a low-mass progenitor of about 10~\Ms\ while on the main sequence. Such stars are predicted to have a very low mass O-core at the time of explosion \citep{rau02, sukhbold_2016}. However, silicate dust is the first to form in the ejecta, while carbon dust forms after 1000 days. The total amount of dust produced, combining the ejecta (0.019 \Ms) and the CSM ($3 \times 10^{-3}$~\Ms), is about 0.022 \Ms. The pre-explosion CSM dust is expected to encounter the forward shock and be destroyed or reprocessed through it. 

\subsection{An Alternate Scenario}

\begin{figure}
\vspace*{0.3cm}
\centering
\includegraphics[width=3.5in]{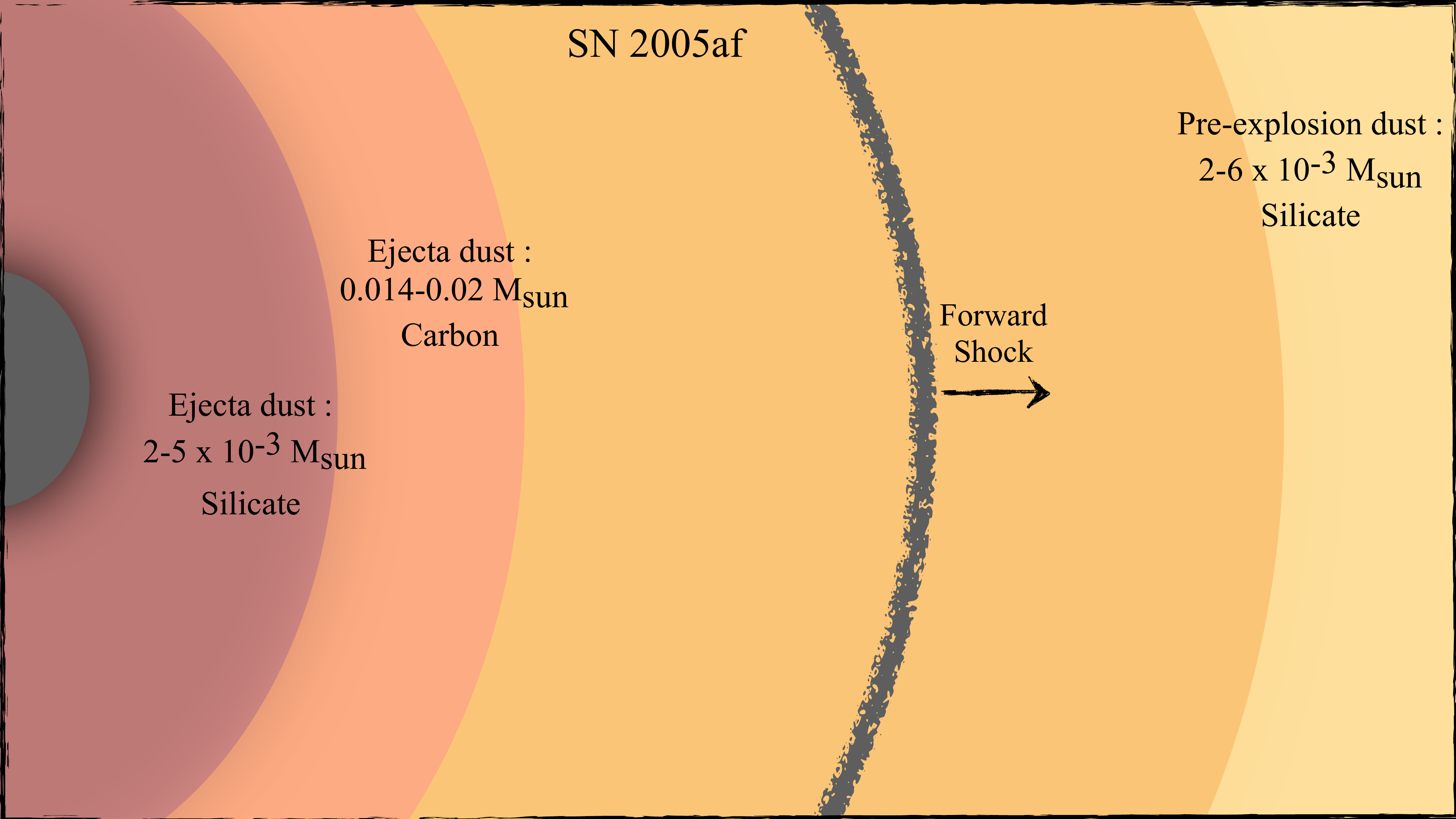}
\caption{\label{fig_results}\footnotesize{A summary of the location, composition, and mass of dust in SN~2005af, after two decades from the explosion, is presented in a schematic diagram (see Section \ref{sec_Spitzer_model} and \ref{sec_Discussion} for details). The inner O/Mg/Si shell, marked in brown, is where silicate dust is formed. The He/C shell, marked in orange, is the site for amorphous carbon dust formation. The dusty region of the CSM, hosting the surviving pre-explosion dust, is marked in yellow. }} 
\end{figure}

When correlating the IR fluxes at different epochs to derive the evolution of dust in an SN, there are several degeneracies in the assumed physical parameters that cannot be resolved solely through the fitting of the data. Specifically, the location of the dust (if it is pre-existing or newly formed after explosion), the shape of the dusty sphere or shell, the relative compositions of different dust types, among other factors, often remain somewhat uncertain. To account for this, there often appear to be multiple scenarios that may all fit the data \citep{shahbandeh_2023, szanna_2024}. In this paper, for SN~2005af, we have presented (summarized in Figure \ref{fig_results}) the most suitable picture which agrees with data as well as the chemical models. 

In an alternative scenario, the IR spectrum at day 629 and the {\it JWST} fluxes at 19~yr can both be fitted using only amorphous carbon dust with mass $\sim$  10$^{-3}$~\Ms (or larger) and 0.015~\Ms, respectively. In this picture, silicate dust is not formed in the ejecta at all. However, the presence of SiO lines on day 272 supports the formation of silicate dust. In addition, detection of Ne and Ar lines indicates that the ejecta should have reasonable O-rich and Si-rich layers where silicate dust is known to form \citep{kotak_2006, crowther_2007, sar13}. Theoretical models suggest that carbon dust formation is delayed owing to the overabundance of He$^+$ ions in the C-rich layers. Based on these conditions, we support the picture where the ejecta of SN~2005af form both silicates and carbon dust, with silicate dust forming earlier than carbon. We acknowledge that in a special case, if the He-rich layer cools down much faster than the inner O-rich part of the ejecta, we may find C-dust to form within the first two years as well.

\section{Discussion}
\label{sec_Discussion}

We report that even after 18~yr post-explosion, SN~2005af remains bright in the mid-IR, with a total IR luminosity $\sim 2.7 \times 10^{4}$ \Ls. The SN ejecta host dust at a temperature of about 100~K at this epoch. In comparison, the mid-IR luminosities of SN~2004et and SN~1980K were found to be $2.2 \times 10^{5}$~\Ls\ and $6.3 \times 10^{4}$~\Ls, respectively \citep{shahbandeh_2023, szanna_2024}. The dust temperatures are also found to be lower in SN~2005af with respect to $\sim 150$~K reported for SN~2004et and SN~1980K. 

As in the cases of SN~2004et and SN~2017eaw \citep{shahbandeh_2023}, we argue that the late-time IR fluxes result from heating of the dust by the power generated by the forward shock and CSM interaction. In the absence of suitable optical data at such late epochs (Keck spectrum at 18~years shown in Figure \ref{fig_lightcurve} does not reveal much about the current state of the SN), we can only calculate a realistic upper limit on the shock power that may heat the dust. Assuming a standard mass-loss rate of $10^{-6}$~\Ms~yr$^{-1}$, wind velocity of 10 km~s$^{-1}$, and shock velocity of 5000~km~s$^{-1}$, the mechanical shock power can be calculated to be $\sim 10^{6}$~\Ls\ \citep{fra14, sarangi2018}. This is sufficient to be assumed as the heating source responsible for the mid-IR fluxes.

There is not much information available about the CSM of SN~2005af. Light-curve models suggest that the progenitor may have lost $\sim 0.33$~\Ms\ of its H-envelope through mass loss in its entire red supergiant phase. There is no significant evidence for the presence of dense CSM anywhere close to the star. Our analysis suggests that there is at least 2~$\times$~10$^{-3}$~\Ms\ of dust in the CSM that has survived the explosion. In addition, there is an upper limit for the distance of the CSM dust, based on the dust temperatures of 180--150~K between days 100--250. The nature of mass loss may also affect the CSM dust, which is not constrained by observations. In a future study, we shall formulate a model for the IR echo from the CSM dust at early times to account for the geometry of the pre-existing dust and the CSM. 

The final mass of dust is between 0.02--0.03 \Ms\ in SN~2005af. This compares well with SN~2004et, where the estimated mass is 0.01--0.05~\Ms\ \citep{shahbandeh_2023}. In SN~1980K, the mass was found to be an order of magnitude smaller \citep{szanna_2024}. Here, our main focus is on the degeneracies between the location of dust and the chemical type of dust. Apart from spatially resolved very nearby objects like SN~1987A, this is the first study that differentiates between SN dust formed prior to the explosion and post-explosion. Moreover, we also, for the first time, present a model of dust formation in a low-mass SN progenitor ($\sim 10$~\Ms at main sequence).


As concluding remarks, we confirm that the metal-rich core of SN ejecta is a very suitable site for dust formation. Based on our dust-formation model results, the dust-to-metal percentage in the ejecta is found to be $\sim 12$\%. Comparing the model with mid-IR fluxes from {\it JWST}, we suggest that carbon and silicate dust components may remain in the ejecta at different temperatures. We report that SN~2005af produced predominantly C-rich dust in the ejecta and O-rich dust in the pre-explosion winds. 

\bigskip
\bigskip

We acknowledge funding from NASA/{\it JWST} grant GO-2666. 
A.S. is thankful for the support by a grant from VILLUM FONDEN, Denmark (PI Jens Hjorth) (project number 16599) and the Department of Science and Technology (DST), Gov. of India. 
S.Z. is supported by the ÚNKP-23-4-SZTE-574 New National Excellence Program of the Ministry for Culture and Innovation from the source of the National Research, Development and Innovation Fund, Hungary.
A.V.F. is grateful for financial support from the Christopher R. Redlich Fund, Arthur Folker, Tom and Dana Grogan, Stan Schiffman, Richard Sesler, and many other donors. 
I.D.L. acknowledges funding from the Belgian Science Policy Office (BELSPO) through the PRODEX project "JWST/MIRI Science exploitation" (C4000142239) and funding from the European Research Council (ERC) under the European Union's Horizon 2020 research and innovation program DustOrigin (ERC-2019-StG-851622).

\bibliographystyle{aasjournal}
\bibliography{Bibliography_sarangi}

\end{document}